\documentclass[runningheads]{llncs}
\usepackage{graphicx}
\usepackage{comment}

\usepackage{array}
\usepackage{xcolor}
\begin{document}
\title{Enterprise Architecture as an Enabler for a Government Business Ecosystem:\\ Experiences from Finland}
\titlerunning{Enterprise Architecture as an Enabler for a Government Business Ecosystem}
%
\author{Reetta Ghezzi\inst{1} 
\and
Taija Kolehmainen\inst{1} 
\and
Manu Setälä\inst{2} 
\and
Tommi Mikkonen\inst{1} 
}
\authorrunning{R. Ghezzi et al.}
%

\institute{University of Jyv{\"a}skyl{\"a}, Jyväskylä, Finland\\
\and Solita, Tampere, Finland \\
\smallskip
\email{reetta.k.ghezzi@jyu.fi, taija.s.kolehmainen@jyu.fi, manu.setala@solita.fi, tommi.j.mikkonen@jyu.fi}
}
\maketitle              
\begin{abstract}
Public sector procurement units in the field of ICT suffer from siloed, application-specific architectures, where each system operates in isolation from others. As a consequence, similar or even identical data is maintained in several different databases hosted by different organizations. Such problems are caused by the lack of standard guidelines and practices that would result in interoperable systems instead of overlapping ones.
In the Finnish public sector, enterprise architecture (EA) is a mandatory requirement so that an ecosystem can be formed to overcome the above problems. However, the adoption rates are low, and the focus is often on technology rather than processes and practices. 
This study investigates the use of EA and its potential in Finnish procurement units through semi-structured interviews. Five procurement units and four vendors participated in the study, and altogether 12 interviews took place.
As a result of the study, a practical implication is establishing decentralized project management practices in procurement units and enhancing leadership to establish a holistic EA. Furthermore, EA maturity evolution increases agility in the procurement unit. 

\end{abstract}

\section{Introduction}

Public organizations follow procurement directives when procuring goods and services, including software. The implementation of procurement directives can vary nationally, and there are no international standards for purchasing software for the public sector. However, most procurement directives aim at ensuring transparency, fairness, and cost-effectiveness in the procurement process. Procurement directive generally includes vendor selection requirements, contract negotiations, and software management. For the latter, enterprise architecture (EA) is a commonly used tool that defines guidelines for how the public organization in question operates and uses IT and data. These guidelines then form the basis for a business ecosystem that delivers services to the public organization.



Ecosystem development is one of the cross-cutting priorities for developing strategic and responsible procurement practices in Finland \cite{keino2022}. By establishing ecosystem thinking, public procurers can pool their resources, relate to innovations taking place in businesses, and maximize their market power and impact.
A business ecosystem is a network of interdependent, loosely interconnected organizations, individuals, and other entities that co-create value \cite{iansiti2004strategy,adner2017ecosystem}, by, for example, distributing goods and services \cite{peltoniemi2004business}.
The business ecosystem concept encompasses the entities that make up a business environment, including suppliers, customers, competitors, regulators, and other stakeholders \cite{iansiti2004strategy}. These ecosystem actors have a specific position in the ecosystem; they are linked to each other and undertake activities to create and capture value in the ecosystem \cite{adner2017ecosystem}. Each component of a business ecosystem affects and is affected by the others, creating a complex web of dependencies \cite{peltoniemi2004business}. For example, changes in one ecosystem component, such as introducing new software, may cause ripples throughout the entire ecosystem and lead to changes in other components. 
Hence, a business ecosystem can be seen as a symbiotic, living organism constantly evolving and adapting to changes in its environment in a robust manner \cite{peltoniemi2004business,iansiti2004strategy}. By examining the relationships and interdependence within the ecosystem, organizations can identify opportunities to respond to challenges and boost their performance.




In this paper, we identify the existence of different ecosystems such as digital ecosystems \cite{briscoe2006digital}, software ecosystems \cite{jansen_cusumano_2013} and digital platform ecosystems \cite{hein2020digital}. We generally concentrate on government business ecosystems where actors interact and transact to co-create value in the context of public procurement.
As a concrete contribution, we present a study that examines the state of EA in the Finnish public sector, and its ability to facilitate a government business ecosystem. In this study, twelve semi-structured interviews are performed with actors that participate in building public sector EA and have a holistic understanding of what could be done to evolve further. 

The rest of this paper is structured as follows. In Section 2, we present the benefits of mature EA compared to what business ecosystem creation demands. In Section 3, the research approach is given, and the research method is described. In Section 4, we present our results. In Section 5, we discuss the results, and in Section 6, we draw some final conclusions.

\section{Background and Motivation}

Characteristics commonly associated with the software include ease of deployment, modifiability, and scalability. The same code can be used in different organizations and different applications. A well-designed approach allows data sharing between other software systems, for instance. Hence, there is no need to re-produce similar software as long as the software components used are generic and reusable instead of monolithic applications. 

In Finland, a certain level of national EA is mandatory, but practical implementations by different actors vary. Fundamentally, with roots in the Open Group Architecture (TOGAF) \cite{cameron_2013}, recommendation JHS-179 \cite{jhs_179} guides how to describe an organization's EA. Unfortunately, while TOGAF is the most widely used EA framework, it has not been thoroughly adopted. In addition, the seminal Zachman framework for enterprise architecture \cite{zachman_1987} is recognized as the foundation of all EA frameworks. This study understands that the Zachman framework is well suited to describe the enterprise architecture of complex and large organizations \cite{zachman_1987}. However, in the public sector context, the Zachman framework is unsuitable for procurement units with little or no IT skills, whereas IT procurement in Finland is commonly carried out by employees whose daily job does not include IT. To this end, we prefer an approach that is intuitively accessible and presents all the interconnections between different roles effortlessly.

Unfortunately, outside the IT domain, procurement unit stakeholder groups fail to adopt EA artifacts in practice \cite{seppanen-keyissues,Nurmi201989}. Public sector software sustainability issues can be overcome with EA, where different services and vendors can quickly deploy and integrate into the ecosystem environment \cite{Setala20213}. Moreover, research performed with 26 practitioners in public agencies reveals that ecosystem thinking in EA software is still missing in practice, even though it is necessary \cite{Nurmi201989}. Furthermore, Nurmi et al. \cite{Nurmi201989} state that public sector EA should utilize the capabilities of the organizations which participate in the ecosystem, develop solutions in co-creation, hold a holistic view over EA, and have need-based EA modeling to enhance ecosystem formation. 

Unfortunately, these viewpoints do not reveal how the public sector and vendors position themselves in the public sector digital ecosystem. An ecosystem, where every piece gives something, may be achieved with services that interact via well-defined APIs but with no direct access to other services \cite{Setala20213}. Techniques in the system need to support systematic and fast development and deployment \cite{Setala20213}. Moreover, public sector software suffers from vendor lock-in, high maintenance costs, and time-consuming and error-prone public tendering. In addition, need-based user utilization, co-creation, holistic view, and organizational capabilities are essential building blocks for public EA \cite{Nurmi201989}. Modular business units \cite{Setala20213}\cite{Ross2006} attached to the organization's core infrastructure help in this regard. 


Improvements in IT efficiencies, such as standardized technology and technology management, lead to increased centralization in management \cite{Ross2006}. The aim is to look forward to shared practices and infrastructure, reduce platforms, and raise cost-effectiveness. The organization's key benefits may require sacrificing some business unit needs \cite{Ross2006,Ross2003,Rakgoale2016214}. Similar findings have been detected among Finnish municipalities. The comparison between the six largest cities in Finland showed that once IT governance becomes centralized and practices somewhat controlled, IT costs and personnel diminish by thirty percent \cite{louhelainen_2013}. 

As a part of digitization, fundamental organizational attitudes need reconsideration, in contrast to traditional processes. When a unit searches for new systems, the negotiating happens among accepted systems and platforms rather than defining a tailored solution and aiming for the best in the markets.  Standardization brings new risks to management; the IT department must be on the nerve to monitor and upgrade the standards. Hence, the complexity of investment decisions rises. The top-management issues haunt hidden behind the problems mentioned above. If the EA lacks top-management sponsorship \cite{seppanen-keyissues,Seppanen2009114,Hjort-Madsen2006}, it is demanding to receive the EA benefits such as cost reduction, IT standardization, process enhancement, and strategic differentiation \cite{Syynimaa2017488}. The lacking leadership hinders EA process adoption. Furthermore, \cite{seppanen-keyissues} recognize that EA practice demands specialized skills and capabilities to manage vast entities. The leader must have leadership and management skills and an understanding of the technical side of the entity. The following list summarises the benefits mature and well-managed EA for an organization:

\begin{itemize}
    \item EA effectively manages IT assets and aligns IT investments and requirements in business \cite{Pour2019189,Bradley201173,Kearns20031,Rakgoale2016214,Ross2006}.
    \item High maturity in EA is a prerequisite for agility in an organization \cite{Bradley201173,Ross2006}.
    \item Ea maturity development enhances the formation of modular business units, where unit managers regain their power by giving them a greater choice to design front-end interfaces \cite{ross_2006}.
    \item Modular business units enable selective standardization by module \cite{ross_2006}, and cost-effective IS replacements \cite{Setala20213}.
    \item IT \cite{ross_2006,louhelainen_2013} and personnel costs diminish \cite{louhelainen_2013}. 
    \item Agility increases through EA, which builds on modular business unit information systems \cite{Setala20213}. 
\end{itemize}  
 
However, EA modeling seems insufficient in terms of digital ecosystem creation. Anwar and Gill \cite{anwar_gill_2019} thoroughly analyzed the seven most common EA frameworks and discovered that the existing frameworks, such as TOGAF, provide tools to support the business and information layers, but not social and professional layers. In this research, we consider these layers to be of utmost significance. Moreover, it seems that existing frameworks could be combined to create a framework to offer a holistic view of EA in digital ecosystem creation \cite{anwar_gill_2019}. 


Maneuvering complex ecosystem interdependencies demands organizations to move towards a more holistic and dynamic mindset, instead of concentrating on controlling the current resources \cite{dattee2018maneuvering}. The ecosystem approach introduces new requirements for structure and functions in value creation, in comparison to, e.g., networks, clusters, and innovation systems. However, understanding the complex ecosystem dynamics and system behavior is challenging \cite{dattee2018maneuvering,basole2015understanding}. 
In this research, we concentrated on evaluating the following characterization of ecosystems: 
\begin{itemize}
    \item Scalability via, e.g., greater integrability and standardization \cite{sklyar2019organizing}.
    \item Adaptivity via, e.g., open and adaptive resource integration \cite{sklyar2019organizing}. 
    \item Shared alignment via, e.g., mutual agreement and compatible incentives \cite{adner2017ecosystem}. 
    \item Dynamic nature via, e.g., improved agility \cite{dattee2018maneuvering}.
    \item Higher interoperability in terms of multilateral connections \cite{adner2017ecosystem}. 
    \item Partnership via, e.g., fostering collaboration and flexibility in control over the ecosystem \cite{iansiti2004strategy}. 
    \item Value co-creation via, e.g., innovation \cite{iansiti2004strategy}. 
    \item Service digitization as it is indispensable for ecosystem creation \cite{sklyar2019organizing}. 
\end{itemize}

We realize that the above-mentioned characterization is not comprehensive, and that it is collected to observe public sector EA and ecosystem initiatives. In this research, we aim at recognizing how the ecosystem-creation inhibitors such as silo structure and rigidness \cite{sklyar2019organizing}, lack of robustness \cite{dattee2018maneuvering}, low need for central control \cite{iansiti2004strategy}, high control over ecosystem \cite{iansiti2004strategy}, and high dominance in value \cite{sklyar2019organizing} present themselves in public sector EA and ecosystem initiatives.


\section{Research Approach}

\textbf{Research Setup and Data Collection}. The participants selected for the study all have experience in public procurement practices and enterprise architecture development in the public sector. The goal was to find which kinds of relationships exist in ICT procurement between procurement units and vendors and how public sector EA guides this process. In some cases, the chosen organizations cooperated with each other or had collaborated previously. The upcoming changes in Finnish public sector infrastructure guide us to examine the state of Finnish public sector EA. 
%
The research question we seek to answer is:
\begin{verse}
    \textit{How does enterprise architecture support digital ecosystem development in the public sector?}
\end{verse}
%
\noindent
Semi-structured interviews were performed between November 2021 and May 2022. The initial literature search and media attention on the Finnish public sector IS project failures  \cite{Kolehmainen_2022,Tivi_2020} presented points to be considered themes in the interviews. These themes were ICT vision, public procurement, financials, IS life cycle, know-how, and commitment. The themes guided the discussions, but the participants were encouraged to contribute what they felt was important. The interview duration varied from 45 min to 63 min. Sometimes intriguing topics need to be discussed more thoroughly. The average time was 55 minutes. Table \ref{tab:interview_data} presents the participant info. 


\begin{table}[tp!]
\begin{center}
\caption{Interview participants.}
\scriptsize
\begin{tabular}{@{}ccccc@{}}
\textbf{Organization} &
  \textbf{Abreviation} &
  \textbf{Position} &
  \textbf{Field} &
  \textbf{Duration} \\ 
Vendor 1          & V1   & Senior Principal   & ICT               & 49 \\
Vendor 2          & V2A  & Head of department & ICT               & 49 \\
Vendor 2          & V2B  & Specialist         & ICT/Procurement & 49 \\
Vendor 3          & V3   & Chief position     & ICT               & 45 \\
Vendor 4          & V4   & Vice President     & ICT/Sales       & 56 \\
Procurement unit 1 & PU1  & Chief position     & ICT               & 47 \\
Procurement unit 2 & PU2A & Manager position   & ICT               & 48 \\
Procurement unit 2 & PU2B & Senior Specialist  & ICT               & 62 \\
Procurement unit 3 &
  PU3A &
 Head of procurement &
 Procurement &
  63 \\
Procurement unit 3 & PU3B & Manager position   & ICT               & 49 \\
Procurement unit 4 & PU4  & Chief position     & ICT               & 58 \\
Procurement unit 5 & PU5  & Manager position   & ICT               & 56
\end{tabular}
\vspace{-6mm}
\end{center}
\label{tab:interview_data}
\end{table}


\textbf{Data Analysis}. All the interviews were recorded and transcribed. The identification data and the repetitions or when the interviewee or interviewer searched for the words were removed. Coding took place in Atlas.ti software's cloud version. 
The approach was inductive, and the phenomena in the data had a guiding role. Hence, the initial coding and theme formation was data-driven, as well as intuitive and reactive \cite{myers2020qualitative}, producing 99 initial codes, and 21 themes. Comparing the themes with literature, Ross et al. \cite{Ross2006} four-stage EA maturity model began to make sense. This resulted in five themes; 1) information system procurement objectives, 2) procurement processes, 3) responsibilities and control, 4) perceptions of the legislative environment, and 5) EA solutions. These themes formulated bundles between the initial themes and codes, and Ross' \cite{ross_2006} stages helped to understand the differences between the organizations. 

However, some phenomena did not directly link to the Ross' \cite{ross_2006} model. For these cases, the ecosystem literature revealed the next steps. To gain a more systematic and structured understanding of the public sector and vendors' position in ecosystems that take place in the context of public procurement, we used a domain-specific modeling language called Ecosystem Governance Compass \cite{sroor2022modeling} to model the ecosystem components, interactions, and dependencies. The language concepts were derived from literature and based on a holistic, dynamic system-based view of collaborative ecosystems \cite{laatikainen_li_abrahamsson_2021}. The language objects were divided into four categories representing different aspects of ecosystem governance: governance, business, technology, and legal and regulatory context. Ecosystem Governance Compass announced places where the EA approach failed to interpret the results, which led to the creation of five additional themes: 6) higher sustainability components, 7) value co-creation, 8) shared objectives, 9) dynamic nature, and 10) holistic view.
These ten themes revealed this research's key findings, where the EA and public sector procedures inhibit or facilitate sustainable ecosystem formation.

\section{Results}

Participants are presented with acronyms to introduce our results, where procurement units are PU1, PU2, PU3, PU4, and PU5. Vendors are V1, V2, V3, and V4. To make a difference between multiple participants from one organization, they are presented with letters A and B, for example, V2A and PU2A.
\subsection{Government business ecosystem inhibitors}

\begin{table}[tp!]
\footnotesize
\caption{Government business ecosystem formation inhibitors.}
\begin{center}
\begin{tabular}{p{0.48\textwidth}|p{0.48\textwidth}}
\textbf{Characteristic}          & Ecosystem Related  Characteristic\\ 
\hline
Most commonly used opportunities in public procurement guide towards a rigid waterfall-like development model. &
Inhibitor for dynamic nature \\
\hline
Actors have no shared alignment & Inhibitor for shared goal and objectives creation \\
\hline
Immature EA and lack of control & Missing collaboration and \\ dynamic control \\ 
\hline
Silo structure & Inhibitor for dynamic, adaptive nature  \\
\hline
Vendor lock-in & Inhibitor for dynamic, adaptive nature \\
\hline
Budgeting IT expenses to the procurement units, the IT department &
  Inhibitor for holistic view \\
\end{tabular}
\end{center}
\vspace{-8mm}
\label{tab:inhibitors}
\end{table}
\textbf{Most commonly used opportunities in public procurement guide towards a stiff waterfall-like development model}.
In this study, public agencies use open, restricted, and competitive negotiated procedures in ICT procurement. Open and restricted procedures are the most common ICT procurement procedures in Finland \cite{pp-tasks_holma}. The competitive negotiated procedure leads to better IS procurement outcomes. In other public procurement procedures, the procurement unit must know precisely what they want and need before the tendering. Furthermore, negotiated procedures without tender hand-in-hand in-house procurement are considered emergency solutions.

\textbf{Actors have no shared alignment}.
Sometimes the actors miss mutual agreement on goals, or their incentives are incompatible. The procurement unit is searching for solutions to fulfill legislative tasks. Vendors are looking for new business opportunities, sales, and good word-of-mouth. PU1 and PU2 understand that interviewing the vendors is essential to know whether the common ground exists, whether the vendor is ambitious to engage in the development process, and whether the view over the issues is holistic. Besides monetary motives to engage in an ecosystem, the incentives should be something else too. However, these incentives are not easily detected in public organizations. Ideally, suppose the consortium of vendors builds the product (identification from one, databases from the other, operational control from the third). In that case, genuine cooperation is created to solve the problem of the procurement unit. 
Procurement units agree that the procurement act sets challenges to forming the above-mentioned coalitions. Tendering is error-prone, time-consuming, and difficult to predict outcomes. Therefore developing a genuine ecosystem-like and sustainable consortium is demanding, if not nearly impossible.
Finally, tailored versus ready-made systems seems to divide opinions among vendors and procurement units. PU1, PU2, and PU3 recon that evaluating the purposefulness of the old processes and ways to work is vital when acquiring new systems to determine if something can be done more efficiently.

\textbf{Immature EA and lack of control, silo structure and vendor lock-in}.
The governance of the public organization has a significant role in committing to the EA decisions. However, some of the interviews reveal that, in many cases, public organizations have immature enterprise architectures and inadequate leadership behind them. 
Public organizations that lack firm leadership to support EA initiatives tend to have a silo structure, where the procurement unit has lots of freedom to tailor solutions that fit one procurement unit. In these cases, the IT department remains in the dark about decision-making and purchasing. Furthermore, these organizations do not have EA units to cross-check the information system's interoperability and compatibility with the existing EA. PU3 has developed its practices and has an EA unit to cross-check the projects, IS, and budget. However, the leadership to put holistic EA thinking into practice is missing. PU3A depicts that every procurement unit leader needs to consider EA in mind, which is troublesome, and the actors are not coordinated optimally. Hence, when the procurement unit purchases a system where compatibility with existing EA is not investigated, problems arise, such as silo architecture \cite{sklyar2019organizing}, vendor lock-in, data integrity, data management, and additional development hours leading to exceeding original budgets, 
to name a few common ones. 

To overcome the data integrity problem, PU3 has determined master systems where the data can be edited. PU3 and PU4 have introduced an incentive to get rid of the solutions that are tailored to one unit, but only those information system purchases that exceed the national thresholds proceed to the EA unit's or project portfolio management's evaluation. In PU5, those information system projects that exceed national thresholds also demand upper-level decision-making. However, no one evaluates the new demand against the existing EA, which has caused a challenging situation in PU5. To this end, PU5 depicts in the interview that:
\begin{quote}
  \textit{"We have 1400 information systems."} 
\end{quote}
\noindent
Without established coherent EA practices, procurement units seem to create disposable EA for IS procurement. In PU5, even that failed. The acquired system in PU5 enables structured documentation and is used  throughout the organization and in similar organizations in the area. However, PU5 has encountered difficulties in it:
\begin{quote}
   \textit{"Two things where it fails; in the tendering phase, the organization's EA and the system's architecture were not evaluated, how they would fit. The second thing is leadership. In large entities, such as this system, the discipline should be in place to guide the development."} 
\end{quote}
\noindent
PU4 describes that sometimes the IT department receives the information from the purchase afterward, even if the organization has set processes to inform the IT department on all IS-related purchases. PU4 does not have decentralized project management practices. Before the purchase, necessarily no-one maps out the budget and personnel resources.
%
Even if the chain of command is not explicitly drafted, the actual purchasing is standardized in all public organizations. In this research, all public organizations have procurement teams or units, where experts help to prepare the procurement and are responsible for the tendering phase. The procurement units provide 
well-prepared procurement practices and tenders. The incentives are to avoid legal issues -- especially the market court -- and to offer vendors equal, non-discriminating tendering processes.  

Vendors depict that resources in public organizations may limit which kind of systems are acquired. Smaller public organizations may not have the resources to go through the heavy public procurement in personnel, competence, and funding. The technology seems to be very flexible, and public organizations can get anything they wish for.
%
V1 expresses concerns when the procurement unit outsources requirement analysis solely to the consultant. The vendor may help the procurement unit with technical requirement analysis, but the needs should emerge from procurement unit functions and objectives. Therefore, V1 is concerned when the procurement unit starts the procurement process with requirement definition before the public procurement. It seems to waste resources, especially in cases when the system itself already exists in the market, but the public sector is not aware of it. In this situation, vendors would only need public organization guidance to understand what exists in their technical field already to avoid going to the path of tailored systems. Hence lack of knowledge of the existing technical field, in terms of compatibility and interoperability, guides vendors to produce tailored solutions if the EA is drafted only for the acquisition in hand. These characteristics describe the inefficient scalability adaptivity in an organization \cite{sklyar2019organizing,Nurmi201989,seppanen-keyissues}. In addition, the environment is rigid and lacks robustness.

\textbf{Budgeting IT expenses to the procurement units rather than to the IT department}. 
Budgeting practices may inhibit coherent EA formation and enhance silo structure. Some public organizations distribute the expenses when the procurement unit administers the funds between its functions. It appears that this is not a viable solution and results in overlapping tasks and IS systems in the organization. There is a low need for interdependent relationships and centralization, which inhibit ecosystem creation \cite{iansiti2004strategy}. 
Furthermore, it seems unthinkable that units which do not hold the competence to evaluate IS-related needs are responsible for IS budget and have the freedom to acquire whatever is wanted under the national threshold. This is the situation in PU3, PU4, and PU5; procurement units control the budget. These units suffer from vendor lock-in and have excessively locally tailored systems. 

All procurement units have legislative tasks that guide service production in society. In Finland, norms such as the public procurement act and procurement directive obligate seeking the most advantageous offer through public procurement. Evaluating the most advantageous offer appear to cause issues for the procurement units. The narrative is apparent between the "old" way of evaluating the most advantageous offer and the "new" way. 

PU3 is incentivized to evaluate the cost and quality of the business operations against the receivable benefits. However, the solutions are not assessed holistically against the EA, and EA is not managed top-down. In addition, different unit leaders are supposed to have a clear understanding of the EA. PU3A sees this as a problem. Some units have a clear picture, others do not, and the top management does not rule or guide them to acquire solutions that serve the whole organization. In PU4 and PU5, the current business objectives are towards reduced IT costs.

\subsection{Government business ecosystem facilitators}

\begin{table}[tp!]
\footnotesize
\caption{Government business ecosystem formation facilitators.}
\begin{center}
\begin{tabular}{p{0.48\textwidth}|p{0.48\textwidth}}
\textbf{Characteristic} &
\textbf{Ecosystem Related Characteristic} \\ 
\hline
Mature EA and sufficient control       & Facilitator for dynamic and adaptive nature \\
\hline
Shared ambition to improve practices 
between the public organization and vendor &
Facilitator for shared goals and objectives \\
\hline
Budgeting IS expenses to IT department & Facilitator for holistic view               \\
\hline
Cooperation with universities &
  Interdependencies between stakeholders, Value co-creation and innovation creation \\ 
\end{tabular}
\end{center}
\label{tab:facilitators}
\vspace{-8mm}
\end{table}

\textbf{Mature EA and sufficient control}.
Sufficient control enables EA practices throughout the organization. Moreover, research by Nurmi et al. \cite{Nurmi201989} states that EA modeling should be need-based. In this research, PU1 and PU2 have top-down support for EA endeavors, which allows a coherent EA landscape. In PU1 and PU2, procurement units cannot purchase anything that suits only one unit's purposes.
Hence, these two viewpoints, need-based EA modeling and top-down support, seem to coexist nicely in PU1 and PU2. In these units, operations guide the needs, and the best practices to execute the solutions are holistically evaluated against EA. PU1 and PU2 seek efficient, predictable, and interoperable systems for their EAs. In addition, the procurement units that have top-down determined EA seem to have more uniform purchasing practices. 
PU1 aims to purchase systems as a service solution (SaaS) to the cloud rather than tailored software.  PU1 depicts that they do not have even one developer in the agency and purchase all the software. PU1 has diminished the number of vendors significantly. At first, PU1 had nearly 100 vendors executing the information systems. Furthermore, many of the solutions had a price tag of just under 60 000€, which is the threshold that demands procurement. PU1 representative thinks these solutions were the result of unplanned spending and panic. In recent history, PU1 has then overcome technology standardization which diminished the number of vendors. PU1 has customized software besides the ready-made solutions, aiming to purchase reusable platforms with modifiable user interfaces. It enables PU1 to have standardized technology and keep the core optimized. PU1 shows minimal data and software duplicity, and the systems interoperate. 
%
%
PU4 depicts that the old ridged systems are replaced gradually with new systems, which creates the grounds for developing data management practices. Here, technology-enabled change is a stepping stone toward standardized technology. 

\textbf{Shared ambition to improve practices and make the change between the public organization and vendor}. V4 has plans to scale the most popular product to the markets in a plug-and-play sort of system because market research shows that it is what procurement units want.
%
V1 is interested in producing better systems that interoperate with local systems, enable standardized working environment units across Finland, and improve working habits. V1 depicts that it is not always easy to measure quality-related improvements, which may not manifest immediately but with time.

\textbf{Budgeting IS expenses to the IT department}.
As mentioned earlier, budgeting practices may inhibit or facilitate ecosystem creation. PU1 and PU2 have centralized IS finance management. The procurement units do not control the IS budget. PU3 is transitioning to centralized IS budget management and revising IS budget management responsibilities as the old IS contracts change to new ones.
In PU1 and PU2, the IT department is the financial gatekeeper and the buyer. If the system wished for is suitable with EA and otherwise advantageous, it proceeds to public procurement. This applies similarly to the IS under the national threshold, even if public procurement is unnecessary. This means, for example, hardware or services under €60k.
Public organizations which realize the benefits of centralizing some functions selectively, also understand that the cost at procurement may be an insufficient metric to evaluate the value generated with EA compatibility, planned lifespan expectancy, improved workflows, and knowledge management. 
PU1 depicts that sometimes the legislative tasks are mandatory but lack business cases. Here, the benefits cannot be measured directly with a cost-benefit analysis. Therefore, during ICT procurement, efficiency might seem ostensible, and the benefits may generate over time indirect ways. PU1 and PU2 determined that whatever is purchased needs to be evaluated and considered throughout. For example, PU2 depicts that a potential vendor lock-in does not matter, if it fits EA and is the best option available to solve the problem organization-wide. In these units, the benefit evaluation reaches from monetary evaluation to non-monetary assessment of the functions.

\textbf{Cooperation with universities}.
Procurement units work with universities in research and development projects. PU3 depicts that the procurement unit may receive something that does not exist yet through these projects. For universities, cooperation offers real-world situations and problems to solve for students. PU3 depicts that: 
\begin{quote}
    \textit{"It was calculated that if one person does the recording work, it will take 5 years. Now we are collaborating with the university to develop a robot and artificial intelligence that can read, interpret and retrieve the right things from the drawings of the built environment and convert them into electronic form."}
\end{quote}
Furthermore, collaboration with universities seems to enhance innovation. This facilitates co-evolving capabilities with actors \cite{moore1993predators} and hence, contribute interdependencies and enhance value co-creation in the ecosystem \cite{Nurmi201989}. 

\section{Discussion}

In this work, 
we used Ecosystem Governance Compass to detect the government business ecosystem facilitators and inhibitors. As the result, we found out that ecosystem thinking is mostly missing from public sector EA and purchasing practices. 
In general, public sector software sustainability seems questionable, since the actors do not have compatible incentives for building up collaboration. In contrast, some public organizations have high-expertise units that form a genuine collaborative web, where every unit works towards similar goals, for example, coherent and efficient EA. However, in some public organizations, the shared goals are not identified \cite{adner2017ecosystem}, and working toward them systematically is missing. Public organizations that have identified the goals can develop solutions in co-creation with different units and vendors, which Nurmi et al. \cite{Nurmi201989} have recognized as vital for public organizations to enable the formation of the digital ecosystem. 


Holistic EA, controlled purchasing, and developing systems iteratively with vendors are signs of adaptivity in this research \cite{sklyar2019organizing}. To consider government business ecosystem formation, we realized that when the procurement units consider the procurement act to offer possibilities in the competitive dialogue and innovation partnership opportunities, these organizations could also selectively standardize \cite{ross_2006} and scale solutions from across the organization \cite{sklyar2019organizing}. 

The government business ecosystem helps to form a holistic view of EA for purchasing and budgeting, creating possibilities to scale solutions, and aiding co-creation and innovation within the ecosystem. Satisfaction towards management increases as the EA maturity evolves. Risk management, IT development time, and strategic business impacts improve, similar to the EA maturity benefits found by Ross \cite{ross_2006}. The organization becomes dynamic. Furthermore, procurement units that have created precise and disciplined EA practices do not waste resources in information system procurement by creating disposable EA. 

In contrast, procurement units that struggle to establish EA also struggle to form a government business ecosystem. These organizations have silo structures \cite{sklyar2019organizing}, where different procurement units can determine which solutions to acquire, and the control is insufficient. Vendor lock-in exists in many places, and public procurement is often seen as a risk of receiving a solution that does not comply with the needs. Furthermore, units with silo structures are missing holistic comprehension of the IT landscape in the organization. The budgeting supports this. The procurement units control the budget, including IS-related purchases, which leads to a situation where the shared incentive to build holistic EA is missing. In this case, the procurement unit purchases and solves problems that concern only one unit.  
%
%
%




Exploring EAs in procurement units reveal that the EA initiatives exist in all participating procurement units, even if they might not be visible in practice. In theory, they exist. Some of the results are similar to Seppänen et al. \cite{seppanen-keyissues}, and Nurmi et al. \cite{Nurmi201989}, who discovered low EA adoption rates in Finnish public sector EA. In this study, procurement units with disciplined decision-making practices are higher in EA maturity. The leadership shows throughout the organization, and the strategy exploits the EA practices and purchases. 

The changes are slow in public sector. Hence, to overcome and dissolve the challenges such as silo structure and vendor lock-in, we trust that the EA approach combined with the ecosystem mindset could help the public organizations to gain a more holistic view of their functions. In particular, modeling tools such as Ecosystem Governance Compass provide an excellent way of describing the formation of a holistic relationship-based ecosystem. Furthermore, Nurmi et al. \cite{Nurmi201989} suggest a centralized EA repository that would update in real-time. This could help national efforts to create a single, interoperable EA. 

\textbf{Threats to validity}.
The research method, semi-structured interviews, allowed the interviewees to depict what was relevant to them. However, this might be a weakness as well \cite{myers2020qualitative}, as the data set was large. Luckily, we had expertise from the University of Jyväskylä to contribute to Ecosystem Governance Compass, which helped us to combine complex phenomena in EA and government business ecosystem creation. 
The data collection and analysis follow Myers \cite{myers2020qualitative} semi-structured interviews and thematic analysis guidelines. Data is collected and analyzed iterative way and rigorously, which makes the study's reliability high. However, the researcher's interpretation may have affected the results because the initial coding was intuitive and interpretive. Myers \cite{myers2020qualitative} depicts that inner validity could be improved with triangulation or multiple researcher evaluation. In this research, the authors collaborated to analyze and discuss the categorizations of the codes. 
The results describe facilitators and inhibitors for the government business ecosystem. Interestingly, the results suggest that EA development in public organizations is at very different stages, which may affect the generalisability of the results. In this study, we do not distinguish EA maturity levels in public organizations.

\section{Conclusion}
In this study, we have analyzed if EA acts as an enabler for a government business system in Finland. 
%
As a tool for analysis, we used Ecosystem Governance Compass to recognize factors that either facilitate or inhibit government business ecosystem creation. As a result, the facilitators are mature EA and sufficient control, shared ambitions, centralized IS budgeting, and cooperation with universities. The inhibitors are the insufficient choice of procurement opportunity, not sharing goals and understanding, immature EA and lack of control, and lack of selective centralization in IS budgeting.  The leadership and top-down support for EA practices are highlighted -- the more mature the EA, the firmer leadership and top-down support. Furthermore, all procurement units in this study have adopted one EA section, standardized purchasing, and use a multi-talented procurement unit or team which prepares the call for tender. However, a hinder to agility lies in the practice before the procurement proposal reaches procurement personnel. Higher EA maturity procurement units have decentralized project management, which is missing from the lower EA maturity procurement units. 

In conclusion, future EA frameworks and practices seem to lean on modular business units in an ecosystemic environment. However, the changes are difficult to implement nationally because each organization acquires services only for itself. However, modeling can imitate the chosen standards, and, with approaches such as openEHR \cite{kalra2005openehr}, 
may be practical to combine accurate modeling and serving user needs in detail. However, more research is needed, because such modeling has scarce scientific literature and empirical results. 


%
%
\bibliographystyle{splncs04}
\bibliography{2309.0015}

\end{document}